\documentclass{article}
\usepackage[margin=1.5in]{geometry}
\usepackage[utf8]{inputenc}
\usepackage{amsmath}
\usepackage{amsfonts}
\usepackage{hyperref}
\newcommand{\footremember}[2]{%
    \footnote{#2}
    \newcounter{#1}
    \setcounter{#1}{\value{footnote}}%
}
 
\providecommand{\keywords}[1]
{
  \small	
  \textbf{\textit{Keywords---}} #1
}

\title{Notes on Generalizing the Maximum Entropy Principle to Uncertain Data}
\author{Kenneth Bogert\footremember{unca}{\href{mailto:kbogert@unca.edu}{kbogert@unca.edu}, University of North Carolina Asheville} }
\date{May 2022}

\begin{document}

\maketitle

\begin{abstract}
The principle of maximum entropy is a broadly applicable technique for computing a distribution with the least amount of information possible constrained to match empirical data, for instance, feature expectations. We seek to generalize this principle to scenarios where the empirical feature expectations cannot be computed because the model variables are only partially observed, which introduces a dependency on the learned model.  Generalizing the principle of latent maximum entropy\cite{wang}, we introduce uncertain maximum entropy and describe an expectation-maximization based solution to approximately solve these problems. We show that our technique additionally generalizes the principle of maximum entropy and discuss a generally applicable regularization technique for adding error terms to feature expectation constraints in the event of limited data. We additionally discuss the use of black box classifiers with our technique, which simplifies the process of utilizing sparse, large data sets.
\end{abstract}

\keywords{entropy, maximum entropy, uncertainty, expectation maximization, partial observability, sparse data, deep learning, classifiers}

\section{Introduction}

The principle of maximum entropy is a technique for finding a distribution over some given elements $X \in \mathbb{X}$ that contains the least amount of information in it while still matching some constraints.  It has existed in various forms since the early 20th century but was formalized by Jaynes~\cite{jaynes1957information} in 1957.  In its commonly encountered form, the constraints consist of matching sufficient statistics, or feature, expectations under the maximum entropy model being learned and those observed empirically. 

However, in many cases the feature expectations are not directly observable.  It could be the case that the model contains hidden variables, that some data is missing or corrupted by noise, or that  $X$ is only partially observable using some type of process or sensor. 

As an example, let us take a simple natural language processing model. Using the principle of maximum entropy, each $X$ will be a word in a vocabulary $\mathbb{X}$, and we wish to form a model that matches the empirical distribution of words in a given document, $\Tilde{Pr}(X)$, according to the expectation of some interesting features $\phi_k(X)$.

However, if the data input into such a model is a voice recording then words are never directly observed.  Instead, we may extract observations $\omega$ from the recording that only partially reveal the word being spoken, for instance, if $\omega$ corresponds to phonemes then $Pr(\omega | X)$, the probability of hearing a phoneme for a given word, will not be deterministic as different dialects and accents pronounce the same word in different ways.  Further, a bad quality voice recording may cause uncertainty in the phoneme being spoken, requiring the use of an even more general $\omega$ to correctly model the data. 

With large amounts of sensor-produced data available and applications~\cite{bogert2016}\cite{10.5555/3535850.3535868}\cite{shahryari2017inverse}\cite{kitani} that may make use of it providing the motivation, we seek to generalize the principle of maximum entropy to scenarios with partial observability of the modeled variables.  

\section{Background}

\subsection{Principle of Maximum Entropy}

Commonly, the principle of maximum entropy is expressed as a non-linear program.

\begin{align}
&  \max \limits_{\Delta} \left( -\sum\nolimits_{X \in \mathbb{X}} Pr(X)~ log~Pr(X) \right )\nonumber\\
&  \mbox{{\bf subject to}} \nonumber\\
& \sum \nolimits_{X \in \mathbb{X}} Pr(X) = 1 \nonumber\\
&  \sum \nolimits_{X \in \mathbb{X}} Pr(X) \phi_k(X)  = \sum \nolimits_{X \in \mathbb{X}} \Tilde{Pr}(X) \phi_k(X)  ~~~~~~ \forall k 
\label{eq:max_ent}
\end{align}

Notably, this program is known to be convex which provides a number of benefits.  Particularly relevant is that we may find a close-form definition of $Pr(X)$, and solving the primal problem's dual is guaranteed to also solve the primal problem~\cite{boyd}.

We begin by finding the Lagrangian relaxation of the program.

\begin{align}
\mathcal{L}(\mathbb{X}, \lambda, \eta) &~=~ -\sum\nolimits_{X \in \mathbb{X}} Pr(X)~ log~Pr(X) + \eta \left ( \sum \nolimits_{X \in \mathbb{X}} Pr(X) - 1 \right ) + \nonumber \\
&\sum\limits_{k=1}^K \lambda_k \left ( \sum \limits_{X \in \mathbb{X}} Pr(X) \phi_k(X) - \sum \limits_{X \in \mathbb{X}} \Tilde{Pr}(X)  ~\phi_k(X) \right )
\label{eq:max_ent_lagrange}
\end{align}

Since the program is convex, the Lagrangian function must be as well.  Therefore, when the Lagrangian's gradient is 0 we have found the global maximum. We now can find the definition of $Pr(X)$:

\begin{align}
{\partial \mathcal{L}(\mathbb{X}, \lambda, \eta) \over \partial Pr(X)} & ~=~ -log Pr(X) - 1 + \eta + \sum\limits_{k=1}^K \lambda_k \phi_k(X) \nonumber \\
0 & ~=~ -log Pr(X) - 1 + \eta + \sum\limits_{k=1}^K \lambda_k \phi_k(X) \nonumber \\
Pr(X) &~=~ {e^{\sum\limits_{k=1}^K \lambda_k \phi_k(X)} \over Z(\lambda)}
\end{align}

Where $Z(\lambda) ~=~ e^{-1}e^\eta ~=~ \sum\limits_{X' \in \mathbb{X}} e^{\sum\limits_{k=1}^K \lambda_k \phi_k(X')}$.
Plugging our definition of $Pr(X)$ back into the Lagrangian, we arrive at the dual.

\begin{equation}
\mathcal{L}_{dual}(\lambda) ~=~ log ~Z(\lambda) - \sum\limits_{k=1}^K \lambda_k \sum \limits_{X \in \mathbb{X}} \Tilde{Pr}(X) \phi_k(X)
\label{eq:max_ent_dual}
\end{equation}

Since the dual is necessarily convex, we find the gradient for use with gradient descent.

\begin{align}
    {{\partial \mathcal{L}_{dual}(\lambda)} \over {\partial \lambda_k}} &~=~ \sum\limits_X Pr(X) \phi_k(X) - \sum \limits_{X} \Tilde{Pr}(X) \phi_k(X) \nonumber \\
\end{align}

Note that any convex optimization technique is a valid alternative to gradient descent.

\subsection{Principle of Latent Maximum Entropy}

First presented by Wang et al.~\cite{wang2001}, the principle of latent maximum entropy generalizes the principle of maximum entropy to models with hidden variables that are never empirically observed.

Split each $X$ into $Y$ and $Z$.  $Y$ is the component of X that is perfectly observed, $Z$ is perfectly un-observed and completes $Y$.  Thus, $X = Y \cup Z$ and $Pr(X) ~=~ Pr(Y, Z)$.  Latent maximum entropy corrects for the hidden portion of $X$ in the empirical data by summing over all $Z \in Z_Y$, which is every way of completing a given $Y$ to arrive at a $X$.

\begin{align}
&  \max \limits_{\Delta} \left( -\sum\nolimits_{X \in \mathbb{X}} Pr(X)~ log~Pr(X) \right )\nonumber\\
&  \mbox{{\bf subject to}} \nonumber\\
& \sum \nolimits_{X \in \mathbb{X}} Pr(X) = 1 \nonumber\\
&  \sum \nolimits_{X \in \mathbb{X}} Pr(X) \phi_k(X)  = \sum \nolimits_{Y \in \mathbb{Y}} \Tilde{Pr}(Y) \sum\limits_{Z \in Z_Y} Pr(Z | Y) \phi_k(X)  ~~~~~~ \forall k 
\label{eq:latent_max_ent}
\end{align}

Since $Pr(Z|Y)$ includes $Pr(X)$, the right side of the constraint contains a dependency on the model being learned, meaning the program is no longer convex and only an approximate solution can be found if we still desire a log-linear model for $Pr(X)$.  This leads to an expectation-maximization approach to find a solution. To our knowledge Wang et al. 2001~\cite{wang2001} is the first to apply EM to the principle of maximum entropy to account for incomplete data.  

The methodology and arguments used in this work is very similar to that used in Wang et al.'s~\cite{wang} and so it will not be duplicated here.  The reader is encouraged, however, to review Wang et al.~\cite{wang} for more background and proofs.

\section{Principle of Uncertain Maximum Entropy}

Assume we want a maximum entropy model of some hidden variables $X \in \mathbb{X}$ given we have observations $\omega \in \Omega$.  Critically, we desire that the model does \emph{NOT} include $\omega$ as the observations themselves will pertain solely to the data gathering technique of the observing entity, not the elements or model being observed. We assume the existence of a static observation function $Pr(\omega | X)$.  Our new non-linear program is:

\begin{align}
&  \max \limits_{\Delta} \left( -\sum\nolimits_{X \in \mathbb{X}} Pr(X)~ log~Pr(X) \right )\nonumber\\
&  \mbox{{\bf subject to}} \nonumber\\
& \sum \nolimits_{X \in \mathbb{X}} Pr(X) = 1 \nonumber\\
&  \sum \nolimits_{X \in \mathbb{X}} Pr(X) \phi_k(X)  = \sum \nolimits_{\omega \in \Omega} \Tilde{Pr}(\omega) \sum \nolimits_X Pr(X | \omega) ~\phi_k(X) ~~~~~~ \forall k 
\label{eq:uncertain_latent_max_ent}
\end{align}

Notice that $Pr(X | \omega) = { Pr(\omega | X) Pr(X) \over Pr(\omega)  }$ and therefore in the infinite limit of data where $\Tilde{Pr}(\omega) = Pr(\omega)$ the constraints are satisfied as:

\begin{align}
 \sum \nolimits_{X \in \mathbb{X}} Pr(X) \phi_k(X)  &= \sum \nolimits_{\omega \in \Omega} Pr(\omega) \sum \nolimits_X Pr(X | \omega) ~\phi_k(X) \nonumber \\
 & = \sum \nolimits_{\omega \in \Omega} Pr(\omega) \sum \nolimits_X { Pr(\omega | X) Pr(X) \over Pr(\omega)  } ~\phi_k(X) \nonumber \\
 & = \sum \nolimits_{\omega \in \Omega} \sum \nolimits_X Pr(\omega | X) Pr(X) ~\phi_k(X) \nonumber \\
 & = \sum \nolimits_X  Pr(X) ~\phi_k(X) \sum \nolimits_{\omega \in \Omega} Pr(\omega | X) \nonumber \\
 & = \sum \nolimits_X  Pr(X) ~\phi_k(X)
\end{align}

To attempt to solve Eq~\ref{eq:uncertain_latent_max_ent}, we first take the Lagrangian.

\begin{align}
\mathcal{L}(\mathbb{X}, \Omega, \lambda, \eta) &~=~ -\sum\nolimits_{X \in \mathbb{X}} Pr(X)~ log~Pr(X) + \eta \left ( \sum \nolimits_{X \in \mathbb{X}} Pr(X) - 1 \right ) + \nonumber \\
&\sum\limits_{k=1}^K \lambda_k \left ( \sum \limits_{X \in \mathbb{X}} Pr(X) \phi_k(X) - \sum \limits_{\omega \in \Omega} \Tilde{Pr}(\omega) \sum \limits_X Pr(X | \omega) ~\phi_k(X) \right ) 
\label{eq:lagrange}
\end{align}

Now we find Lagrangian's gradient so that we can set it to zero and attempt to solve for $Pr(X)$.

\begin{align}
{\partial \mathcal{L}(\mathbb{X}, \Omega, \lambda, \eta)} \over {\partial Pr(X)} & ~=~ -log Pr(X) - 1 + \eta  ~+ \nonumber \\
& \sum\limits_{k=1}^K \lambda_k \left (  \phi_k(X) -  \sum \limits_{\omega \in \Omega} \Tilde{Pr}(\omega) \left ( \phi_k(X)  {{ Pr(\omega | X) Pr(\omega) - Pr(\omega | X) ^2 Pr(X)}  \over { Pr(\omega)^2}} \right ) \right ) \nonumber\\
& ~=~-log Pr(X) - 1 + \eta + \sum\limits_{k=1}^K \lambda_k \phi_k(X) \nonumber \\
& - \sum\limits_{k=1}^K \lambda_k \sum \limits_{\omega \in \Omega} \Tilde{Pr}(\omega) \left ( \phi_k(X)  {{ Pr(\omega | X) Pr(\omega) - Pr(\omega | X) ^2 Pr(X)}  \over { Pr(\omega)^2}} \right )
\label{eq:lagrange_gradient}
\end{align}

Unfortunately, the existence of $Pr(X | \omega)$ on the right side of the constraints causes the derivative to be non-linear in $Pr(X)$.  Instead, we will approximate $Pr(X)$ to be log-linear. In other words:

\begin{align}
{\partial \mathcal{L}(\mathbb{X}, \Omega, \lambda, \eta)} \over {\partial Pr(X)} & ~\approx~ -log Pr(X) - 1 + \eta + \sum\limits_{k=1}^K \lambda_k \phi_k(X) \nonumber \\
0 & ~\approx~ -log Pr(X) - 1 + \eta + \sum\limits_{k=1}^K \lambda_k \phi_k(X) \nonumber \\
Pr(X) &~\approx~ {e^{\sum\limits_{k=1}^K \lambda_k \phi_k(X)} \over {Z(\lambda)}}
\end{align}

Now we plug our approximation back into the Lagrangian to arrive at an approximate Dual:

\begin{equation}
\mathcal{L}_{dual}(\lambda) ~\approx~ log ~Z(\lambda) - \sum\limits_{k=1}^K \lambda_k \sum \limits_{\omega \in \Omega} \Tilde{Pr}(\omega) \sum \limits_X Pr(X | \omega) \phi_k(X)
\label{eq:dual}
\end{equation}

We would now try to find the dual's gradient and use it to minimize the dual.  Unfortunately the presence of $Pr(X|\omega)$ still admits no closed form solution in general.  We will instead have to employ another technique to minimize it.


\subsection{Expectation Maximization}

Start with the log likelihood of all the observations:

\begin{align}
L(\lambda) & ~=~ \sum\limits_{\omega \in \Omega} \Tilde{Pr}(\omega) ~log~ Pr_\lambda(\omega) \nonumber \\
& ~=~ \sum\limits_{\omega \in \Omega} \Tilde{Pr}(\omega) ~log~ \sum\limits_{X \in \mathbb{X}} Pr_\lambda(\omega, X) \nonumber\\
& ~=~ \sum\limits_{\omega \in \Omega} \Tilde{Pr}(\omega) ~log~ \sum\limits_{X \in \mathbb{X}} { Pr_\lambda(\omega, X) \over Pr_{\lambda'} (X | \omega)} Pr_{\lambda'} (X | \omega) \nonumber\\
& ~\geq~ \sum\limits_{\omega \in \Omega} \Tilde{Pr}(\omega) \sum\limits_{X \in \mathbb{X}} Pr_{\lambda'} (X | \omega) ~log~ { Pr_\lambda(\omega, X) \over Pr_{\lambda'}(X | \omega) } \nonumber\\
& ~=~ \sum\limits_{\omega \in \Omega} \Tilde{Pr}(\omega) \sum\limits_{X \in \mathbb{X}} Pr_{\lambda'} (X | \omega) ~log~ Pr_\lambda(\omega, X) - \sum\limits_{\omega \in \Omega} \Tilde{Pr}(\omega) \sum\limits_{X \in \mathbb{X}} Pr_{\lambda'} (X | \omega) ~log~  Pr_{\lambda'}(X | \omega) \nonumber\\
& ~=~ \sum\limits_{\omega \in \Omega} \Tilde{Pr}(\omega) \sum\limits_{X \in \mathbb{X}} Pr_{\lambda'} (X | \omega) ~log~ Pr_\lambda(\omega| X) Pr_\lambda(X) + H(\lambda') \nonumber\\
& ~=~ \sum\limits_{\omega \in \Omega} \Tilde{Pr}(\omega) \sum\limits_{X \in \mathbb{X}} Pr_{\lambda'} (X | \omega) ~log~ Pr_\lambda(\omega| X) + \sum\limits_{\omega \in \Omega} \Tilde{Pr}(\omega) \sum\limits_{X \in \mathbb{X}} Pr_{\lambda'} (X | \omega) ~log~ Pr_\lambda(X) + H(\lambda') \nonumber \\
\label{eq:EM_U_star}
& ~=~ \sum\limits_{\omega \in \Omega} \Tilde{Pr}(\omega) \sum\limits_{X \in \mathbb{X}} Pr_{\lambda'} (X | \omega) ~log~ Pr(\omega| X) + Q(\lambda, \lambda') + H(\lambda')\\
& ~=~ U^*(\lambda') + Q(\lambda, \lambda') + H(\lambda')
\end{align}

Eq~\ref{eq:EM_U_star} follows because $Pr(\omega | X)$ is the observation function which does not depend upon $\lambda$.  This leaves $Q(\lambda, \lambda')$ as the only function which depends upon $\lambda$.  The EM algorithm proceeds by maximizing $Q$, and upon convergence $\lambda = \lambda'$, at which time the likelihood of the data is at a local maximum.

$H(\lambda')$ is the conditional entropy on the latent variables, and $U^*(\lambda')$ is the expected log observations, which due the the observations not being included in the model only impacts the overall data likelihood, but not the model solution.

We now plug in a log-linear model for $Pr(X)$ to $Q(\lambda, \lambda')$:

\begin{align}
Q(\lambda, \lambda') & ~=~ \sum\limits_{\omega \in \Omega} \Tilde{Pr}(\omega) \sum\limits_{X \in \mathbb{X}} Pr_{\lambda'} (X | \omega) ~log~ Pr_\lambda(X) \nonumber \\
& ~=~ \sum\limits_{\omega \in \Omega} \Tilde{Pr}(\omega) \sum\limits_{X \in \mathbb{X}} Pr_{\lambda'} (X | \omega) \left (  \sum\limits_{k=1}^K \lambda_k \phi_k(X) - ~log~Z(\lambda) \right ) \nonumber \\
& ~=~ - log~Z(\lambda) + \sum\limits_{k=1}^K \lambda_k \sum\limits_{\omega \in \Omega} \Tilde{Pr}(\omega) \sum\limits_{X \in \mathbb{X}} Pr_{\lambda'} (X | \omega)  \phi_k(X) 
\label{eq:em_finished}
\end{align}

Notice that Eq.~\ref{eq:em_finished} is similar to Eq.~\ref{eq:dual}. One important difference is that Eq.~\ref{eq:em_finished} is easier to solve, as $Pr(X |\omega)$ depends on $\lambda'$ and not $\lambda$.  In fact, maximizing $Q(\lambda, \lambda')$ is equivalent to solving the following program:

\begin{align}
&  \max \limits_{\Delta} \left( -\sum\nolimits_{X \in \mathbb{X}} Pr_\lambda(X)~ log~Pr_\lambda(X) \right )\nonumber\\
&  \mbox{{\bf subject to}} \nonumber\\
& \sum \nolimits_{X \in \mathbb{X}} Pr_\lambda(X) = 1 \nonumber\\
&  \sum \nolimits_{X \in \mathbb{X}} Pr_\lambda(X) \phi_k(X)  = \sum \nolimits_{\omega \in \Omega} \Tilde{Pr}(\omega) \sum \nolimits_X Pr_{\lambda'}(X | \omega) ~\phi_k(X) ~~~~~~ \forall k 
\label{eq:uncertain_latent_max_ent_em}
\end{align}

which equals Eq.\ref{eq:uncertain_latent_max_ent} at convergence.  We now arrive at the following Expectation-Maximization algorithm:\\\\
\emph{Initial Start:} Randomly initialize $\lambda'$\\
\emph{E Step:}  Using  $\lambda'$, compute $\hat{\phi}_k ~=~ \sum \nolimits_{\omega \in \Omega} \Tilde{Pr}(\omega) \sum \nolimits_X Pr_{\lambda'}(X | \omega) ~\phi_k(X)$\\
\emph{M Step:} Solve the following convex program to arrive at a new $\lambda$:
\begin{align}
&  \max \limits_{\Delta} \left( -\sum\nolimits_{X \in \mathbb{X}} Pr_\lambda(X)~ log~Pr_\lambda(X) \right )\nonumber\\
&  \mbox{{\bf subject to}} \nonumber\\
& \sum \nolimits_{X \in \mathbb{X}} Pr_\lambda(X) = 1 \nonumber\\
&  \sum \nolimits_{X \in \mathbb{X}} Pr_\lambda(X) \phi_k(X)  = \hat{\phi}_k ~~~~~~ \forall k 
\label{eq:em_m_step}
\end{align}
Then set $\lambda' = \lambda$\\
\emph{Repeat:} Until $\lambda$ converges 

\section{Specializations}

Here we demonstrate that the principle of uncertain maximum entropy generalizes both the principle of maximum entropy and the principle of latent maximum entropy\cite{wang} by showing that we recover these earlier methods when certain specific conditions are met.\\\\

\subsection{Principle of Maximum Entropy:}\label{sec:standardmaxent}
  We recover the  Principle of Maximum Entropy if $Pr(X | \omega) \in \{0, 1\} ~\forall~ X, \omega$ and $\exists ~\omega ~\ni~ Pr(X | \omega) ~=~ 1 ~\forall~ X$.  In other words, each $\omega$ specifies a single $X$ deterministically.  Note that the reverse is not necessarily true, $Pr(\omega | X)$ need only be deterministic if $|\Omega| = |\mathbb{X}|$.  However, for a given $X$ specified by a given $\omega$:

\begin{align}
    Pr(X | \omega) &~=~ { Pr(\omega | X) Pr(X) \over Pr(\omega)  } \nonumber \\
    1 &~=~ { Pr(\omega | X) Pr(X) \over Pr(\omega)  } \nonumber \\
    Pr(\omega) &~=~ Pr(\omega | X) Pr(X) \nonumber
\end{align}

Therefore, in Eq~\ref{eq:lagrange_gradient} (the Lagrangian's gradient), the final term is always zero and we find $Pr(X)$ is log linear (without approximation), and Eq.~\ref{eq:dual} is exact.  Furthermore, as $Pr(X | \omega)$ is unaffected by $\lambda$ the gradient of Eq.~\ref{eq:dual} may now be found as:

\begin{equation}
    {{\partial \mathcal{L}_{dual}(\lambda)} \over {\partial \lambda_k}} ~=~ \sum\limits_X Pr(X) \phi_k(X) - \sum \limits_{\omega \in \Omega} \Tilde{Pr}(\omega) \sum \limits_X Pr(X | \omega) \phi_k(X)
\end{equation}

Thus, we do not need to use EM to solve this problem, and we have arrived at a principle of Maximum Entropy solution.  To see terms that exactly match, let $|\Omega| = |\mathbb{X}|$, then $\Tilde{Pr}(\Omega) = \Tilde{Pr}(X)$ and $\sum\limits_X Pr(X|\omega) = \sum\limits_{X'} Pr(X'|X)$. 

\begin{align}
\mathcal{L}_{dual}(\lambda) &~=~ log Z(\lambda) - \sum\limits_{k=1}^K \lambda_k \sum \limits_{X \in \mathbb{X}} \Tilde{Pr}(X)  \sum \limits_{X'} Pr(X' | X)  \phi_k(X) \nonumber \\
    &~=~ log Z(\lambda) - \sum\limits_{k=1}^K \lambda_k \sum \limits_{X \in \mathbb{X}} \Tilde{Pr}(X) \phi_k(X) \nonumber \\
\end{align}
\\
\subsection{Principle of Latent Maximum Entropy:}\cite{wang} This technique breaks up $X$ into two components, $Y$ which is perfectly observed, and $Z$ which is missing (perfectly un-observed) and $X = Y \cup Z$.  To show that Maximum Entropy with Uncertain Observations generalizes latent maximum entropy, we must show a reduction of the right side of the main constraint to $\sum\limits_{Y} \Tilde{Pr}(Y) \sum\limits_{Z \in Z_Y} Pr(Z | Y) \phi_k(X)$ when $Pr(Y | \omega) \in \{0, 1\} ~\forall~ Y, \omega$ and $\exists ~\omega ~\ni~ Pr(Y | \omega) ~=~ 1 ~\forall~ Y$.  In other words, each $\omega$ specifies a single $Y$ deterministically.  Note that the reverse is not necessarily true, $Pr(\omega | Y)$ need only be deterministic if $|\Omega| = |\mathbb{Y}|$. 

Using this definition,
\begin{align}
    Pr(Y | \omega) &~=~ { Pr(\omega | Y) Pr(Y) \over Pr(\omega)  } \nonumber \\
    1 &~=~ { Pr(\omega | Y) Pr(Y) \over Pr(\omega)  } \nonumber \\
    Pr(\omega) &~=~ Pr(\omega | Y) Pr(Y) \nonumber
\end{align}

Now note that $Pr(X) = Pr(Y, Z)$, we have

\begin{align}
    Pr(X | \omega) &~=~ { Pr(\omega | Y, Z) Pr(Y, Z) \over Pr(\omega)  } \nonumber \\
    &~=~ { Pr(\omega | Y, Z) Pr(Z | Y) Pr(Y) \over Pr(\omega | Y) Pr(Y) } \nonumber \\
    &~=~ {Pr(\omega | Y, Z) Pr(Z | Y) \over Pr(\omega | Y) } \nonumber \\
    &~=~ {Pr(\omega | Y) Pr(Z | Y) \over Pr(\omega | Y) } \label{eq:lmesimpl} \\
    &~=~ Pr(Z | Y)    
\end{align}

Since $Z$ is perfectly unobserved, $Pr(\omega | Y, Z) = Pr(\omega | Y)$ on eq~\ref{eq:lmesimpl}.  To match terms exactly, let $|\Omega| = |Y|$, then $\tilde{Pr}(\Omega) = \tilde{Pr}(Y)$. Notice, whenever $Pr(Y | \omega) = 0, Pr(X | \omega) = 0$.  Therefore we may ignore the summation term in these cases, and only consider $Z \in Z_Y$:\\
$\sum\limits_\omega \Tilde{Pr}(\omega) \sum\limits_{X} Pr(Z | Y) \phi_k(X) ~=~ \sum\limits_Y \Tilde{Pr}(Y) \sum\limits_{Z \in Z_Y} Pr(Z | Y) \phi_k(X)$

\section{Large, sparse observation sets}

The desire to automate inference has driven the use of extremely large, sparse datasets produced by machine sensors as the input into various learning models.  Often techniques such as deep neural networks are used to transform the sparse dataset into dense data, perhaps the model elements directly.  These techniques may be trained by making use of supervised learning on a subset of the available data that has been manually labeled.  As the output learned model may be a black box, no human discernible observation features may be available for examination for use in uMaxEnt. Here we extend the principle of uncertain maximum entropy to these black box scenarios.

Suppose we have an enormous, sparse dataset $\mathbb{R}$ from which samples $r$ are produced from the model element's true observation features, ie. $Pr(r | \omega)$ ($r$ stands for \textit{raw data}).  These $r$ samples are what is received by the observer, as $\Omega$ is unknown to the observer, and may be thought of as encoded (possibly partially) into $r$.  For example, if the observer is using a RGB camera, $\omega$ may be a 3D mesh describing the full visual representation of a particular $X$ and $r$ is a 2D RGB image of the mesh.

Suppose we are given a set of samples from $\mathbb{R}$ labeled by a human.  To increase generality, we allow the labels to be from a different, though related, set than $\mathbb{X}$.  Let $\xi \in \Xi$ be these labels and define a function $d(X, \xi) \xrightarrow{} \{0, 1\}$ that is 1 when a given $X$ maps to a given $\xi$.  For simplicity of argument we restrict $d$ such that each $X$ maps to only one $\xi$ deterministically.  Note that the opposite need not be true, one $\xi$ may map probabilistically to many $X$. Extension to more general configurations is straightforward and only involves modifying $Pr(\xi | X)$ appropriately, we will not discuss this further here.

Now, we may employ some method to classify all received $r$ into $\xi$.  Let $F(r) \xrightarrow{} \xi$ be the function learned by this method, and let us further assume this method comes with statistical performance metrics such as precision and recall.  Then, we use $F$ to classify all available sparse data into $\tilde{Pr}(\xi)$ and our new uMaxEnt constraints for this scenario are:

\begin{align}
\sum \limits_{X \in \mathbb{X}} Pr(X) \phi_k(X)  &= \sum\limits_{\xi \in \Xi} \tilde{Pr}(\xi) \sum \limits_{X} Pr(X | \xi) \phi_k(X)
\end{align}

Where $Pr(X | \xi) ~=~ {Pr(\xi | X) Pr(X) \over Pr(\xi) }$, and $Pr(\xi | X)$ is the probability that $F$ outputs $\xi$ when the true, underlying model element present is $X$.  This is given by the method's performance metrics, though possibly with appropriate modification to account for the difference between $\Xi$ and $\mathbb{X}$. Note that in the event the classification method used is perfect, these new constraints revert to either latent maximum entropy (when $\Xi \subset \mathbb{X}$) or standard maximum entropy (when $\Xi = \mathbb{X}$).  

\subsection{Uncertain classification}

Suppose that the classification method used cannot be certain as to which $\xi$ should be output for a given $r$ and instead produces a distribution over $\xi$, $Pr(\xi | r)$.  This provides only partial information of which $\xi$ is present, but has an advantage in that the method encodes the accuracy of its output into the output distribution itself.  This in turn greatly simplifies $Pr(\xi | X) = d(X, \xi)$.

Our first attempt at using this distribution may be to find the expected $\xi$ as follows:

\begin{align}
\sum \limits_{X \in \mathbb{X}} Pr(X) \phi_k(X)  &= \sum \limits_{r} \tilde{Pr} (r) \sum\limits_{\xi \in \Xi} Pr(\xi | r) \sum \limits_{X} Pr( X | \xi) \phi_k(X)
\end{align}

However, this faces an issue as the distribution $Pr(\xi | r)$ is produced using the training data set, and not the specific dataset under consideration.  To see this, suppose the method used is parameterized with $\theta$, and we provide an infinite amount of data such that $\tilde{Pr}(r) = Pr(r)$:

\begin{align}
& \sum \limits_{r} Pr(r) \sum\limits_{\xi \in \Xi} Pr_\theta(\xi | r) \sum \limits_{X} Pr(X | \xi) \phi_k(X) \nonumber \\
= & \sum \limits_{r} \sum\limits_{\xi \in \Xi}  Pr_\theta(\xi , r) \sum \limits_{X} Pr(X | \xi) \phi_k(X) \nonumber \\
= & \sum\limits_{\xi \in \Xi}  Pr_\theta(\xi) \sum \limits_{X} Pr(X | \xi) \phi_k(X) \nonumber \\
\neq &  \sum\limits_{\xi \in \Xi}  \sum \limits_{X} Pr(X, \xi) \phi_k(X) \\
\end{align}

Because  the \textbf{training} distribution over $\xi$, $Pr_\theta (\xi)$, can vary dramatically from the target distribution $Pr(\xi)$, we cannot guarantee that this method produces an effective approximation.  For instance, suppose in the training set the distribution over $\xi$ was deliberately chosen to be uniform in order to prevent bias in the learning, whereas this distribution is highly unlikely to be the correct one in an inference task.

To correct for this, we examine $Pr_\theta(\xi | r)$ using Baye's law.  Note that even if the method used does not allow for these components to be separated as shown they still must be represented in some capacity in order to produce a valid distribution.

\begin{align}
Pr_\theta(\xi | r) &~=~ {Pr_\theta(r | \xi) Pr_\theta(\xi) \over Pr_\theta(r) }
\end{align}

We note that $Pr_\theta(r)$ is a normalizer, and so we target $Pr_\theta(\xi)$ and replace it with $Pr(\xi)$, then normalize to obtain an updated distribution. 

\begin{align}
Pr_\theta(\xi | r){Pr(\xi) \over Pr_\theta(\xi)} &~=~ {\nu~ Pr_\theta(r | \xi) Pr_\theta(\xi) Pr(\xi)  \over Pr_\theta(\xi)} \nonumber \\
&~=~ \nu'~ Pr_\theta(r | \xi) Pr(\xi) \nonumber \\
&~=~ {Pr_\theta(r | \xi) Pr(\xi) \over \sum\limits_{\xi'} Pr_\theta(r | \xi') Pr(\xi') }
\end{align}

Where $\nu$ and $\nu'$ are normalizers, and differ as we require renormalization after the correction.  Now, notice $Pr(r) = \sum\limits_\xi Pr(r | \xi) Pr(\xi) $,  which differs from the normalizer above only in the term $Pr_\theta(r | \xi)$, which is the observation model being learned by the classification technique.  This term is expected to approximate the true observation model as closely as possible, as that is the whole purpose of employing the technique! 

So we arrive at, in the case of infinite data:

\begin{align}
\sum \limits_{X \in \mathbb{X}} Pr(X) \phi_k(X)  &= \sum \limits_{r} Pr(r) \sum\limits_{\xi \in \Xi} { Pr_\theta(\xi | r) Pr(\xi) \over Pr_\theta(\xi)} \sum \limits_{X} Pr( X | \xi) \phi_k(X) \nonumber \\
&~\approx~ \sum \limits_{r} \sum\limits_{\xi \in \Xi}  Pr_\theta(r | \xi) Pr(\xi) \sum \limits_{X} Pr( X | \xi) \phi_k(X) \nonumber \\
&~=~  \sum\limits_{\xi \in \Xi}  Pr(\xi) \sum \limits_{X} Pr( X | \xi) \phi_k(X) \sum \limits_{r} Pr_\theta(r | \xi) \nonumber \\
&~=~ \sum\limits_{\xi \in \Xi}  Pr(\xi) \sum \limits_{X} Pr( X | \xi) \phi_k(X) \nonumber \\
&~=~ \sum\limits_{\xi \in \Xi} \sum \limits_{X} Pr( X , \xi) \phi_k(X) \nonumber \\
&~=~ \sum \limits_{X} Pr( X ) \phi_k(X) 
\end{align}

The quality of the approximation is now controlled by the quality of the classification technique, this a desirable trait as the classification technique's quality is controlled by the engineers building or training it.

This variant of uMaxEnt incorporates $Pr(X)$ twice since $Pr(\xi) = \sum\limits_X Pr(\xi | X) Pr(X)$.  Notice that even in the event that $\Xi = \mathbb{X}$ we still have a uMaxEnt problem, due to the presence of this second $Pr(X)$ and ultimately caused by the uncertainty in $Pr_\theta(\xi | r)$.

\section{Discussion}

The principle of uncertain maximum entropy makes explicit that the choice of model influences results by including $Pr(X)$ in the empirical side of the constraints.  In cases where uncertainty exists in $Pr(X | \omega)$ this technique ensures a model is found that is consistent with the available information and not over-committed to the specific observations received, as would be the case with ignoring the uncertainty and using the principle of maximum entropy, perhaps by taking the expectation, mean, or maximum $X$ given $\omega$.

Another benefit of this technique is existing $Pr(X)$ priors may be used in the first E step of the expectation-maximization algorithm, somewhat similar to how it is done with Bayesian methods, as opposed to uninformative priors.  This can help bias the results to reflect earlier experiences that cannot, for whatever reason, be included in $\tilde{Pr}(\omega)$.






\bibliographystyle{abbrv}
\bibliography{main}

\begin{thebibliography}{1}

\bibitem{10.5555/3535850.3535868}
K.~Bogert and P.~Doshi.
\newblock A hierarchical bayesian process for inverse rl in
  partially-controlled environments.
\newblock In {\em Proceedings of the 21st International Conference on
  Autonomous Agents and Multiagent Systems}, AAMAS '22, page 145–153,
  Richland, SC, 2022. International Foundation for Autonomous Agents and
  Multiagent Systems.

\bibitem{bogert2016}
K.~Bogert, J.~F.-S. Lin, P.~Doshi, and D.~Kulic.
\newblock Expectation-maximization for inverse reinforcement learning with
  hidden data.
\newblock In {\em Proceedings of the 2016 International Conference on
  Autonomous Agents \& Multiagent Systems}, pages 1034--1042, 2016.

\bibitem{boyd}
S.~Boyd and L.~Vandenberghe.
\newblock {\em {Convex Optimization}}.
\newblock 2002.

\bibitem{jaynes1957information}
E.~T. Jaynes.
\newblock Information theory and statistical mechanics.
\newblock {\em Physical review}, 106(4):620, 1957.

\bibitem{kitani}
K.~M. Kitani, B.~D. Ziebart, J.~A. Bagnell, and M.~Hebert.
\newblock {Activity forecasting}.
\newblock {\em Computer Vision–ECCV}, pages 201--214, 2012.

\bibitem{shahryari2017inverse}
S.~Shahryari and P.~Doshi.
\newblock Inverse reinforcement learning under noisy observations.
\newblock {\em arXiv preprint arXiv:1710.10116}, 2017.

\bibitem{wang2001}
S.~Wang, R.~Rosenfeld, and Y.~Zhao.
\newblock Latent maximum entropy principle for statistical language modeling.
\newblock In {\em IEEE Workshop on Automatic Speech Recognition and
  Understanding, 2001. ASRU'01.}, pages 182--185. IEEE, 2001.

\bibitem{wang}
S.~Wang, D.~Schuurmans, and Y.~Zhao.
\newblock The latent maximum entropy principle.
\newblock {\em ACM Trans. Knowl. Discov. Data}, 6(2), July 2012.

\end{thebibliography}

\end{document}